\documentclass[final,3p]{elsarticle}

\usepackage{amsfonts,latexsym,eucal,amsmath,amsthm,amssymb} 
\usepackage{graphicx}
\usepackage{psfrag}
\usepackage{color}
\usepackage{amsbsy}     
\usepackage{lineno}
\usepackage{flushend}
\usepackage{cuted}

\def\a{\alpha}

\def\dd{\mbox{d}}

\def\ve{\varepsilon}

\def\x{{\bf x}}

\def\n{{\bf n}}

\def\z{{\bf z}}

\journal{Journal of Theoretical Biology}
\begin{document}
\begin{frontmatter}

\title{Fixation times in differentiation and evolution in the presence of
  bottlenecks, deserts, and oases}

\author{Tom Chou$^{1,2}$\corref{cor1}\fnref{}}
\ead{tomchou@ucla.edu}
\cortext[cor1]{Corresponding author}
\fntext[]{Tel: 310-206-2787, Fax: 310-825-8685}


\author{Yu Wang$^{3}$}
\address{$^{1}$Dept. of Biomathematics, UCLA, Los Angeles, CA 90095-1766, USA \\
$^{2}$Dept. of  Mathematics, UCLA, Los Angeles, CA 90095-1555, USA \\
$^{3}$Institute for Information and System Sciences, Xi'an Jiaotong University,
Xi'an, China}





\begin{abstract}
Cellular differentiation and evolution are stochastic processes that can
involve multiple types (or states) of particles moving on a complex,
high-dimensional state-space or ``fitness'' landscape. Cells of each specific type
can thus be quantified by their population at a corresponding node
within a network of states. Their dynamics across the state-space
network involve genotypic or phenotypic transitions that can occur
upon cell division, such as during symmetric or asymmetric cell
differentiation, or upon spontaneous mutation. Waiting times between
transitions can be nonexponentially distributed and reflect {\it
  e.g.,} the cell cycle. Here, we use a multi-type branching processes 
to study first passage time statistics for a single cell to appear in
a specific state. We present results for a sequential evolutionary
process in which $L$ successive transitions propel a population from a
``wild-type'' state to a given ``terminally differentiated,''
``resistant,'' or ``cancerous'' state.  Analytic and numeric
results are also found for first passage times across an evolutionary
chain containing a node with increased death or proliferation
rate, representing a desert/bottleneck or an oasis. Processes involving cell
proliferation are shown to be ``nonlinear'' (even though mean-field
equations for the expected particle numbers are linear) resulting in
first passage time statistics that depend on the position of the
bottleneck or oasis.  Our results highlight the sensitivity of
stochastic measures to cell division fate and quantify the limitations
of using certain approximations and assumptions (such as
fixed-population and mean-field assumptions) in evaluating fixation
times.
%
\end{abstract}


\end{frontmatter}


\section{Introduction}
 
Stochastic models of populations have long been applied to biological
processes such as stem cell dynamics \cite{JAGER2009,ROSHAN2014},
tumorigenesis
\cite{PORTIER1996,PORTIER2000,BELLACOSA2003,SPENCER2006,ATTOL2010,ANTAL2011},
cellular aging \cite{FRANK2005}, and organismal evolution
\cite{LJSALLEN,ANTAL2010}.  In such applications, one is often
interested in the statistics of the time it takes for members of a
population to first arrive at a specific ``absorbing'' state.  Such a
state may represent, for example, a high fitness phenotype that
eventually takes over the entire population.

\begin{figure}[t]
\begin{center}
\includegraphics[width=3.6in]{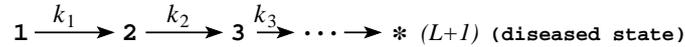}
\caption{Multistage model in disease progression. When multiple steps
  are before the system reaches diseased state $L+1$, an $L-$fold
  convolution of the state-dependent individual waiting time
  distributions provides the overall waiting time distribution and
  the survival probability against disease.}
\label{STAGES}
\end{center}
\end{figure}

A classic biomedical application of first passage times of a
single conserved entity arises in models of cancer progression that
attempt to describe the survival probability of patients as a function
of time after initial diagnosis or treatment. In the Knudsen
hypothesis of cancer progression (illustrated in Fig.~\ref{STAGES})
\cite{KAPLAN,MOOL}, an individual acquires a certain number of
sequential mutations or ``hits'' before acquiring cancer
\cite{KNUDSEN,FITZGERALD1983,BELLACOSA2003}.  
If multiple rare transitions are required before onset of disease, we
can define the probability of transition from state $\ell$ to state
$\ell+1$ in time $\dd t$ as $w_{\ell}(t)\dd t$. The overall
waiting-time distribution to first arrive at the diseased state
$\ell=L+1$ is a convolution of all the $w_{\ell}(t)$ and can be easily
expressed using Laplace transforms: $\tilde{W}(s) =
\prod_{\ell=1}^{L}\tilde{w}_{i}(s)$.  Since each mutation is
considered rare, the event times of each mutation are exponentially
distributed. If all transitions occur at the same rate, $k_{0} = k_{1}
= \ldots = k$, $w_{\ell}(t) = ke^{-kt}$, and $\tilde{W}(s) =
k^{L}/(s+k)^{L}$. The inverse Laplace transform then gives
\cite{FLOYD2010}

\begin{equation}
W(t)\dd t = k{(kt)^{L-1} e^{-kt} \over (L-1)!}\dd t.
\end{equation}
This expression assumes that all the transition rates are 
equally rate-limiting. If $kt \ll 1$, the survival probability against disease onset
is approximately

\begin{equation}
S(t) = 1-\int_{0}^{t}W(t')\dd t' = {\Gamma(L,kt)\over (L-1)!} \approx 1-{(kt)^{L}\over L!},
\end{equation}
If sufficiently accurate fitting of this expression to measured $S(t)$
can be performed, the number of mutations, or ``hits'' $L$ before onset of
cancer can be inferred. Using this Knudsen hypothesis \cite{KNUDSEN}, typical
cancers have yielded $L\sim 4-15$ or higher
\cite{RIEKER2008,BEERENWINKEL2007}.

Such studies implicitly assume a ``single-particle'' picture of a
conserved random walker that eventually reaches a target.  On a
cellular level this picture is appropriate for a single immortal and
nonproliferating cell that successively acquires different mutations.
Estimates and scaling relationships of first passage times of
conserved particles on complex networks have been developed in more
general contexts\cite{KAHNG2012,AGLIARI2008}. Analogous results have
been developed for a fixed multiple number of noninteracting particles
\cite{WEISS1980}. Inverse problems (similar to the inference of the
number of mutations in Knudsen's hypothesis) have also been recently
explored. Li, Kolomeisky, and Valleriani \cite{KOLO2014} considered
how first passage times of a conserved random walker can be used to
estimate the shortest paths to the absorbing site, even for
nonexponentially distributed waiting times between jumps within the
network.  First passage times of Brownian motion and random walks have
also been used to infer properties of continuous energy landscapes
\cite{BAL2003,FOKPRS}.

If a network is finite, and all nodes are connected, conserved
particles will always arrive at an absorbing state and the survival
probability $S(t\to\infty) \to 0$.  However, in the presence of other
pathways for particle annihilation, the absorbing site may never be
reached.  Additional particles need to be continuously injected into
the network in order for one of them to eventually arrive, with
certainty, at a specific absorbing state \cite{FPTREVIEW}. Alternative
annihilation pathways and immigration lift the fixed population
constraint and is an essential feature in cell and population biology.


Canonical Wright-Fisher and Moran models of evolution consider a
population of organisms distributed between two states
\cite{LJSALLEN}.  Evolution across multiple states or fitness levels
have also been explored in models of stochastic tunneling
\cite{ISAWA2004,CHAO2005,FISHER2009}.  Many of these models impose a
fixed mean populations and do not resolve the possible microscopic
transitions an organism can take during the evolution process.  These
differences in the ``microscopic'' mechanisms of evolution are
especially distinguishable in cell biology, in which changes in
genotype or phenotype can arise spontaneously in an individual cell,
or from symmetric or asymmetric replication. Different cell fates are
clearly important in the context of stem cell differentiation and
cancer \cite{ANTAL2010,ANTAL2011,ROSHAN2014,MCHALE2014}.  Moreover,
due to cell death, cell populations can have high turnover within the
timescale of their evolution. Therefore, the total instantaneous
population need not be fixed, even if the ensemble-averaged population
remains constant. We shall see that the different transitions inherent
in cellular differentiation and evolution, as well as fluctuations in
population, can qualitatively affect fixation
times.

We begin by considering a whole population of cells or ``particles''
in a network.  Fixation in this context will be defined by a single
cell or particle first arriving at an absorbing node.  Absorbing nodes can
represent, for example, terminally differentiated, fully
drug-resistant, or highly fit, fully cancerous states.  We first treat
only a noninteracting population and temporarily neglect any
regulation or population constraint such as carrying capacity.  The
analysis is simplified when the total population is unconstrained;
however, we will extend mathematical framework in order to resolve the
effects of different types of allowed transitions. To describe the
evolution of a whole population of cells and their arrival times to
the absorbing nodes, we exploit a multi-type Bellman-Harris branching
process that allows for general distributions of waiting times between
transition events \cite{NEY1972,LJSALLEN,FOK_PROOF}.  Our approach is
related the analysis of Portier, Sherman, and Kopp-Schneider
\cite{PORTIER2000} and the simulations of Sherman and Portier
\cite{PORTIER1996}, but we provide numerical, asymptotic, and exact
mean-field results to illustrate the effects of order-dependent
transition rates. New approximations for analyzing processes
constrained by carrying capacity are also developed.

In the next section, for completeness, we present the continuous-time
semi-Markov multi-type branching formalism and derive the equations
obeyed by the probability generating functions for particle numbers at
each node in the network.  The corresponding equations for the
survival probabilities are then derived. By further assuming
exponentially distributed waiting times and a sequential evolution
model, we explicitly derive the matrix Riccati equation governing the
evolution of survival probabilities in the presence of immigration. In
the Results, we present analytic, asymptotic, and numerical results
for survival probabilities and mean first passage times. Effects of
the probabilities of the different cellular transitions on our results
are explored. A breakdown of mean-field theories of survival
probabilities (even when particles are noninteracting) is
described. Effects of heterogeneity in the transition rates are
discussed in the context of evolutionary oases and bottlenecks.  The
conditions under which the order of the transition rates along the
evolutionary chain can affect the survival probabilities and first
passage times are investigated.  Finally, we summarize our results,
discuss related biological applications, and describe extensions and
future directions.

\section{Mathematical Model}
Here, we describe in detail a stochastic multi-type population in the
presence of immigration.  The general framework is presented before
restricting ourselves to exponentially distributed inter-transition
times and sequential evolution for a more detailed analysis.

\subsection{Multi-type Branching Process}


Our analysis of the problem is most efficiently performed using an
age-dependent multi-type branching process where a parent cell of type
$k$ waits a time $\tau$ before dividing into a number of cells of
possibly different types.  Cells with different numbers of mutations,
or at different stages of differentiation, can have different
distributions of waiting times before proliferation. Moreover, each
cell type, upon proliferation, can yield different numbers of new
cells. 
%
%
In the analysis of this multi-type branching process,
we employ the probability generating function (pgf)

\begin{equation}
F_{k}({\bf z};t) = 
\sum_{n_{1}=0}^{\infty}\cdots\sum_{n_{L+1}=0}^{\infty}
P_{k}({\bf n};t) z_{1}^{n_{1}}\cdots z_{L+1}^{n_{L+1}},
\end{equation}
in which ${\bf z} = (z_{1}, z_{2}, \ldots, z_{L}, z_{L+1})$ and ${\bf
n} = (n_{1}, n_{2}, \ldots, n_{L}, n_{L+1})$.  $P_{k}({\bf n};t)$ is
the probability at time $t$ the entire population contains $n_{j}$
cells of type $j$, given that the system started at $t=0$ with a
single cell of type $k$. We assume that all daughter cells proliferate
independently and that each branching event of a single cell of type
$k$ yields $m_{1}, m_{2},
\ldots, m_{L+1}$ cells of type $1,2,\ldots, L+1$ with probability
$a^{(k)}(m_{1}, m_{2}, \ldots, m_{L+1})\equiv a^{(k)}({\bf m})$.  

What equation of evolution does $F_{k}({\bf z};t)$ obey?  For
notational simplicity, it is easiest to first consider a
single-species branching process described by the simple pgf $F(z,t)$
that corresponds to $P(n,t\vert 1,0)$, the probability of $n$
particles at time $t$, given a single parent particle at $t=0$.  If we
now define $F(z,t\vert \tau)$ as the generating function of the
process conditioned on the original parent particle having first
``branched'' between $\tau$ and $\tau+\dd \tau$, we write the recursion


\begin{equation}
F(z,t\vert \tau) = \bigg\{ \begin{array}{ll}
z, & t< \tau \\[13pt]
A[F(z,t-\tau)], & t \geq \tau,\end{array}
\label{BH0}
\end{equation}
where 

\begin{equation}
A[z] = \sum_{m=0}^{\infty} a(m) z^{m}
\label{A1}
\end{equation}
defines the probability $a(m)$ that a particle splits into $m$
identical particles upon branching.  We now average
Eq.~(\ref{BH0}) over the distribution of waiting times between 
branching events, $g(\tau)$, to find

\begin{equation}
\begin{array}{rl}
F(z,t) & \displaystyle \equiv \int_{0}^{\infty}F(z,t \vert \tau)g(\tau)\dd \tau \\[13pt]
\: & \displaystyle = z \int_{t}^{\infty}\!\!\!g(\tau)\dd \tau + 
\int_{0}^{t}\!A[F(z,t-\tau)]g(\tau)\dd \tau.
\end{array}
\label{BELLMANHARRIS}
\end{equation}
This Bellman-Harris branching process \cite{NEY1972,FOK_PROOF} is
defined by two parameter functions, $a(m)$, the vector of progeny
number probabilities, and $g(\tau)$, the probability
density function (pdf) for waiting times between branching events for
each particle.  Given a single-particle initial condition, $F(z,0)=z$
and Eq.~\ref{BELLMANHARRIS} can be solved to find a $F(z,t)$, from which
$P(n,t\vert 1,0)$ can be generated.

For our multi-state model, we simply generalize Eq.~\ref{BELLMANHARRIS}
to a multi-type process, where particles at different states constitute
different types. The vector of progeny probabilities $a(m)$ now becomes a matrix
$a^{(k)}({\bf m})$ coupling the birth of different types of particles from a parent
particle of state $k$. Thus,

\begin{equation}
A_{k}[{\bf z}] \equiv
\sum_{m_{1}=0}^{\infty}\!\!\cdots\!\!\!\sum_{m_{L+1}=0}^{\infty}
a^{(k)}({\bf m})z_{1}^{m_{1}}\cdots
z_{L+1}^{m_{L+1}}
\end{equation}
is the pgf of the progeny number distribution matrix associated with each
branching event. The relationship for the multi-type pgf becomes

\begin{equation}
F_{k}({\bf z};t) = z_{k}\int_{t}^{\infty}\!\!\!g_{k}(\tau)\dd\tau
+ \int_{0}^{t}\!\!A_{k}\left[{\bf F}({\bf z};t-\tau)\right] g_{k}(\tau)\dd \tau,
\label{F}
\end{equation}
where $g_{k}(\tau)\dd \tau$ is the probability that a particle of type
$k$ branches between time $\tau$ and $\tau +\dd \tau$ after it was
created.

Consider $S_{k}(t) = F_{k}(z_{j\neq L+1}=1, z_{L+1}\to 0^{+}; t)$, the
probability of not having formed a cell of type $L+1$ up to time $t$
given one initial parent cell in node $k$ at time $t=0$.  Setting
$z_{j\neq L+1}=1$ in Eq.~\ref{F}, we find

\begin{equation}
S_{j \neq L+1}(t) = \int_{t}^{\infty}\!\!\!g_{j}(\tau)\dd\tau + \int_{0}^{t}
\!\!A_{j}\left[{\bf S}(t-\tau)\right]g_{j}(\tau)\dd\tau,
\label{SJ0}
\end{equation}
where ${\bf S} = \{S_{j\neq L+1}\}$ is the vector of survival
probabilities initiated by a single cell in state $j$. Since $L+1$ is
defined as an absorbing state, we are interested in the first time a
particle first arrives at node $L+1$.  Therefore, by setting $A_{L+1}=0$, we
allow particles to only accumulate in state $L+1$, and define the survival
probability $S_{L+1}(t)=F_{L+1}(z_{i\neq L+1}=1,z_{L+1}=0)=0$. This
``boundary condition'' in the starting positions, along with the
initial conditions $S_{j\neq L+1}(t=0)$, completely defines the
problem for ${\bf S}(t)$.

Note that our model neglects particle-particle interactions
and that the transition probabilities $a^{(k)}({\bf m})$ do not depend on
the number of particles in the network. Therefore, all initial
particles behave independently and the survival probability associated
with a system initiated with $N$ cells at node $i=1$ is simply
$\Sigma(t) \equiv [S_{1}(t)]^{N}$. Provided that no particles leave the
network other than through state $L+1$, $S_{k}(t\to 0) = o(t^{-1})$,
the {\it mean} first arrival time $T = \int_{0}^{\infty} \Sigma(t)\dd
t$ is well-defined. However, if the particle dynamics include death,
there can be extinction before node $L+1$ is reached, and the mean
arrival time $T$ will diverge. In this case, a more useful measure of
the speed of evolution would be the mean arrival time conditioned on
arrival at $L+1$ \cite{FPTREVIEW}.

A process that ensures arrival to the final state $L+1$ is injection
of particles from an external source. We can extend the branching
process formulation to include immigration of parent particles into
the system \cite{JAGERS1968,SHONKWILER1980}. Suppose that particles of
type $i$ are injected into the system with inter-injection times
distributed according to $h_{i}(\tau)$. Upon assuming an initially
{\it empty} network, the pgf for the total particle numbers resulting
from independently injecting type $i$ particles is thus \cite{JAGERS1968,SHONKWILER1980}

\begin{equation}
\Phi_{i}({\bf z};t) = \int_{t}^{\infty}\!\!\!\!\!h_{i}(\tau)\dd\tau + 
\int_{0}^{t}\!\!\Phi_{i}({\bf z};t-\tau) B_{i}
[F_{i}({\bf z};t-\tau)]h_{i}(\tau)\dd\tau, 
\label{PHI}
\end{equation}
where $B_{i}[z_{i}] = \sum_{n_{i}=0}^{\infty}
b_{i}(n_{i})z_{i}^{n_{i}}$ is the pgf constructed from the probability
$b_{i}(n_{i})$ that $n_{i}$ particles are simultaneously injected into
state $i$ during each immigration event. For example, if particles are
injected only three-at-a-time into node $i$, $b_{i}(n_{i}) =
\delta_{n_{i},3}$.  In the cellular biology setting, immigration into
the $i^{\rm th}$ state can arise from spontaneous mutation or from
mutations acquired during replication of an ``external'' (not included
in the states $k$) wild-type cell.  Therefore, $B_{i}[F_{i}] =
b_{i}(1)F_{i} + b_{i}(2) F_{i}^{2}$, where $b_{i}(1)$ and $b_{i}(2)$
are the probabilities that during each event, one and two cells
immigrate into state $i$, respectively.  Since these are the only
allowed mechanisms of cellular immigration, $b_{i}(1)+b_{i}(2)=1$.  In
the presence of immigration into all possible stages, the pgf of the
total particle number is thus $\Psi({\bf z};t) =
\prod_{i=1}^{L}\Phi_{i}({\bf z};t)$.

Upon using Eqs.~\ref{F} and \ref{PHI} to find $\Psi({\bf z};t)$, one
can construct quantities such as the expected number of cells of type
$k$, $\langle n_{k}(t)\rangle = (\partial \Psi({\bf z};t)/\partial
z_{k})\vert_{{\bf z}={\bf 1}}$, and the probability that no cells have
yet reached the fully mutated state $i=L+1$: $\Sigma(t) =
\Psi(z_{j\leq L}=1, z_{L+1}\to 0; t)$.  Without loss of generality, we
henceforth restrict our analysis to immigration only into node
$i=1$. This limit can be explicitly constructed by letting the times
between consecutive immigration into stages $i > 1$ diverge.  For
example, if $h_{i\neq 1}(\tau) = \lim_{T_{i}\to
  \infty}\delta(\tau-T_{i})$, Eq.~\ref{PHI} then yields $\Phi_{i\neq 1} \to
1$ and $\Psi({\bf z};t) = \Phi_{1}({\bf z};t)$.

When $i=1$, Eq.~\ref{PHI} shows that the overall survival
probability $\Sigma(t)$ in the presence of cell immigration obeys

\begin{equation}
\Sigma(t) = \int_{t}^{\infty}\!\!\!h_{1}(\tau)\dd\tau + 
\int_{0}^{t}\!\Sigma(t-\tau)B_{1}[S_{1}(t-\tau)]h_{1}(\tau)\dd\tau,
\label{SIGMA0}
\end{equation}
By solving Eqs.~\ref{SJ0} for $S_{1}(t)$ and using the result in
Eq.~\ref{SIGMA0}, we can find the overall survival probability of an
initially empty network after cells begin to immigrate into state
$i=1$.  Since cells are not conserved (in particular, they can die),
$S_{j\neq L+1}(t\to \infty)$ need not vanish. However, provided
particle injection into state $i=1$ persists, the absorbing state will
eventually be reached with certainty and $\Sigma(t\to\infty) \to
0$. Depending on the immigration frequency and number of imported
particles per injection event, reaching the terminal state may be
rate-limited by either the internal dynamics defined by $a^{(k)}({\bf
  m})$ and $g(\tau)$, or by immigration described by $b_{i}(n_{i})$
and $h_{i}(\tau)$. Finally, the {\it mean} first passage time (MFPT)
can be calculated from \cite{FPTREVIEW,REDNERFPT}

\begin{equation}
T = \int_{0}^{\infty} \Sigma(t)\dd t.
\label{T}
\end{equation}

\subsection{Exponentially distributed sequential processes}

Our results can be simplified if branching and immigration times
are exponentially distributed, $g_{j}(\tau) =
\lambda_{j}e^{-\lambda_{j}\tau}$ and $h_{1}(\tau) =
\beta_{1}e^{-\beta_{1}\tau}$. After some algebra, Eqs.~\ref{SJ0} and
\ref{SIGMA0} become

\begin{equation}
{\dd S_{k}(t)\over \dd t} =\lambda_{k} A_{k}[{\bf S}(t)] -
\lambda_{k}S_{k}(t),
\label{MATRIXS}
\end{equation}

\begin{equation}
{\dd \Sigma(t)\over \dd t} =
-\beta_{1}\left(1-B_{1}[S_{1}(t)]\right)\Sigma(t).
\end{equation}
Thus, the survival probability can be explicitly expressed as 

\begin{equation}
\Sigma(t) =
\exp\left[-\beta_{1}\int_{0}^{t}\left(1-B_{1}[S_{1}(t')]\right)\dd
  t'\right],
\label{SIGMAEXP}
\end{equation}
where $S_{1}(t)$ is found from solving Eq.~\ref{MATRIXS}. 

\begin{figure}[t]
\begin{center}
\includegraphics[width=3.6in]{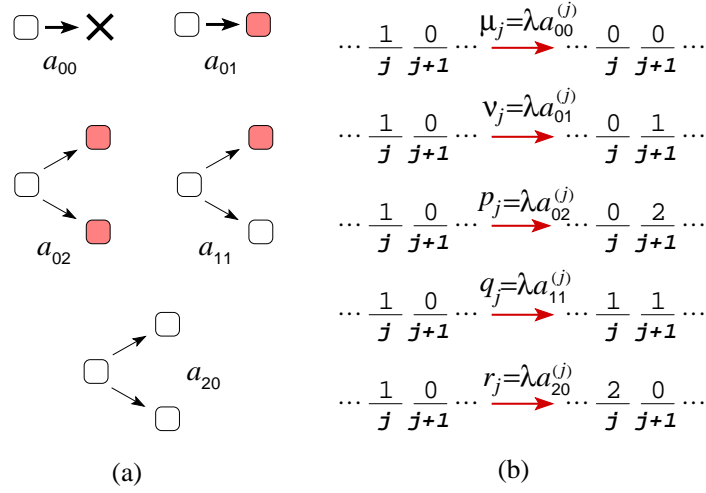}
\caption{(a) The five possible transitions of a single cell at an
  initial stage (white) and their probabilities $a_{mn}$.  Dividing
  cells can produce daughters at a more differentiated or mutated
  stage (red). Since these are the only possible steps, $a_{00}+
  a_{01}+a_{11}+a_{20}+a_{02} = 1$. Note that in general time between
  transitions can be age-dependent and do not have to be exponentially
  distributed in time. (b) When inter-transition times are
  exponentially distributed, the rates of each process can be defined
  in terms of the branching rate and the branching probabilities
  $a_{mn}^{(k)}$ at each node $k$.}
\label{TRANSITIONS}
\end{center}
\end{figure}

The analysis can be further simplified by assuming a sequential
evolution processes where each division by a cell can yield only
daughter cells of the same type or of an incrementally more
differentiated (or mutated) type.  In other words, when a type $k$
cell attempts to proliferate, either death occurs, or daughters of
only type $k$ and/or $k+1$ are produced. Consequently, $a^{(k)}({\bf
  m})=0$ for any $m_{j} > 0$ when $j\neq k, k+1$. Therefore, $F_{k+1}$
in Eq.~\ref{F} is coupled to $F_{k}$ through the integrand
$A_{k}[F_{1}, F_{2},\ldots, F_{L+1}]$, and one must solve for all
$F_{j}$. To be explicit, if the only possible transitions are those
depicted in Fig.~\ref{TRANSITIONS}(a), we find

\begin{equation}
A_{k}[{\bf z}] = a_{00}^{(k)} + a_{01}^{(k)} z_{k+1} +
a_{11}^{(k)}z_{k}z_{k+1} + a_{20}^{(k)}z_{k}^{2} +
a_{02}^{(k)}z_{k+1}^{2}.
\end{equation}
In the context of cell biology, the probabilities $a_{00}, a_{01},
a_{02}, a_{11}$ and $a_{20}$ shown in Fig.~\ref{TRANSITIONS} represent
death, somatic mutation, symmetric differentiation, asymmetric
differentiation, and replication after each attempt at cell division.
For $g_{j}(\tau) = \lambda_{j}e^{-\lambda_{j}\tau}$ the individual
rates of these processes are given by $\mu_{k} =
\lambda_{k}a^{(k)}_{00}, \nu_{k} \equiv \lambda_{k}a^{(k)}_{01}, r_{k}
\equiv \lambda_{k}a^{(k)}_{20}, q_{k}\equiv \lambda_{k}a^{(k)}_{11}$,
and $p_{k} = \lambda_{k}a^{(k)}_{02}$, and shown in
Fig.~\ref{TRANSITIONS}(b). Similarly, we define $\alpha_{1} =
\beta_{1} b_{1}(1)$ and $\alpha_{2} = \beta_{1}b_{1}(2)$ as the rates
of injecting a single particle and double particle into state $i=1$,
respectively.  The values $\mu_{k}, \nu_{k}, p_{k}, q_{k}$, and
$r_{k}$ correspond to {\it rates} of death, somatic mutation, symmetric
differentiation, asymmetric differentiation, and symmetric
replication, respectively, of cells in state $k$.

\begin{figure}[t]
\begin{center}
\includegraphics[width=3.6in]{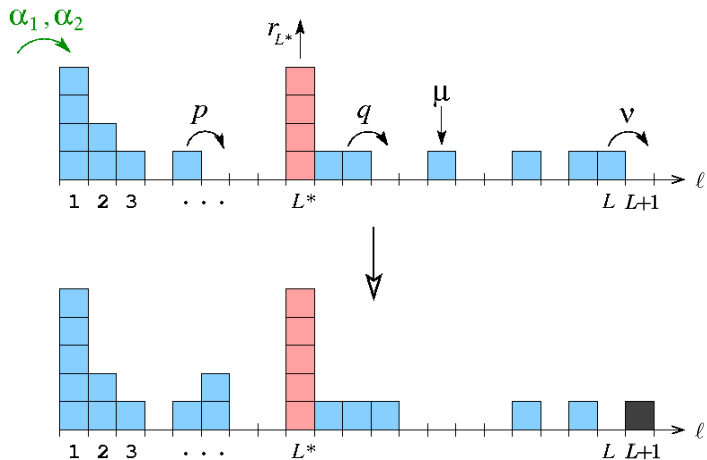}
\caption{A sequential evolution model. The possible transitions are
  labeled. Heterogeneities in the transition rates along the sequence
  can be easily incorporated in our computation. Localized
  heterogeneities ({\it e.g.}, at site $L^{*}$) can be used to model
  oases or bottlenecks. Our analysis focusses on the first arrival
  time to state $L+1$.}
\label{ZRP}
\end{center}
\end{figure}

A sequential evolution model can thus be constructed by assigning a
set of transition probabilities at each successive cell state, or node, as shown in
Fig.~\ref{ZRP}.
Eq.~\ref{MATRIXS} for $S_{k}(t)$ and the associated initial condition
thus reduces to 

\begin{equation}
{\dd S_{k}(t)\over \dd t} = \mu_{k} + \nu_{k}S_{k+1} + r_{k}S_{k}^{2} +q_{k}S_{k}S_{k+1}
+ p_{k} S_{k+1}^{2}-\lambda_{k}S_{k}
\label{SK}
\end{equation}
and $S_{k\leq L}(0) = 1$, $S_{L+1}(t)=0$.


\vspace{5mm}

\section{Results}

In this section, we present both analytic and numeric results for
$S_{k}(t)$, $\Sigma(t)$, and the MFPTs $T$ for sequential, exponentially
distributed processes described by Figs.~\ref{TRANSITIONS} and \ref{ZRP}. We
discuss their properties as functions of transition rates and system
size, and compare these results with those obtained from the simplest
mean-field approximations.
\vspace{2mm}

\subsection{Linear processes}

For ``linear'' dynamics, defined by $p_{k}=q_{k}=r_{k}=0$,
Eqs.~\ref{SK} are linear and can be solved exactly using Laplace
transforms:

\begin{equation}
\tilde{S}_{k}(s) = {1\over s}\left[1-\prod_{i=k}^{L}{\nu_{i} \over (s+\lambda_{i})}\right].
\label{SS}
\end{equation}
This result explicitly shows that $\tilde{S}_{1}(s)$, and hence
$\Sigma(t)$ is invariant with respect to the order of
$\mu_{i}+\nu_{i}=\lambda_{i}$. Therefore, heterogeneity in the
transition rates of this linear Poisson process does not influence the
first passage times to the absorbing state. Similarly, the survival
probability for a sequential process with general waiting time
distribution $g_{k}(\tau)$ can be found from solving Eq.~\ref{SJ0} to
find
$\tilde{S}_{1}(s)=s^{-1}[1-\prod_{i=1}^{L}a_{01}^{(i)}\tilde{g}_{i}(s)]$,
which is also clearly independent of the order of the 
transitions.

Eq.~\ref{SS} can be inverted to obtain explicit expressions for
$S_{k}(t)$. $S_{1}(t)$ can be then used in Eq.~\ref{SIGMAEXP} to obtain
the full survival probability $\Sigma(t)$, and ultimately the MFPT
using Eq.~\ref{T}. For uniform $\lambda_{k} =\lambda$, Eq.~\ref{SS} 
simplifies to

\begin{equation}
S_{k}(t) = 1-\left({\nu \over \lambda}\right)^{L-k+1}\left[1-{\Gamma(L-k+1,\lambda t) \over
\Gamma(L-k+1)}\right],
\end{equation}
which is equivalent to the survival probability of a zero-range
process with death \cite{SHA10}.  

If there is no immigration nor death ($\mu = 0$ and $\lambda = \nu$),
the process is analogous to an irreversible multi-step Moran process in
which a parent cell immediately dies after producing one mutated/evolved/differentiated
daughter cell. The conservation of particles means that eventual
arrival to any connected node $L+1$ is certain. For an initial
condition of $N$ particles in node $k=1$, the mean time for a first
cell to arrive at the terminal state $L+1$ is

\vspace{-3mm}
\begin{equation}
\begin{array}{rl}
T & \displaystyle = \nu^{-1}\int_{0}^{\infty}
\left({\Gamma(L,y)\over \Gamma(L)}\right)^{N}\dd y \\[13pt]
\: &   \displaystyle \sim {1 \over \nu}
\left[{(L-1)\Gamma(L)\over N-1}\right]^{1/L}\!\!\!\!\!\!, 
\quad
{1\over L}\left[{(L-1)\Gamma(L)\over N-1}\right]^{1/L}\!\!\!\!\!\ll 1.
\end{array}
\end{equation}

\begin{figure}[h!]
\begin{center}
\includegraphics[width=3.6in]{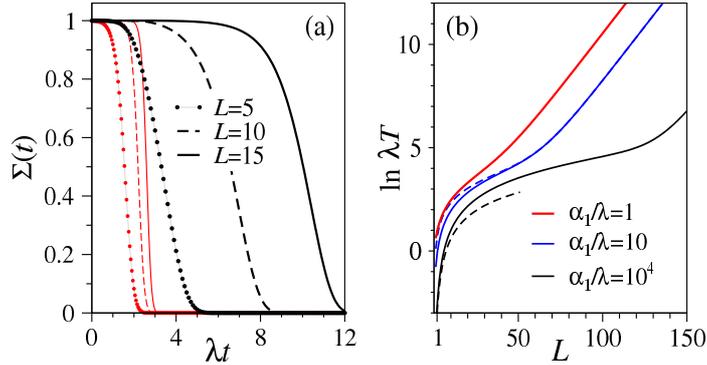}
\vspace{-1mm}
\caption{(a) Survival probabilities for $L=5,10,15$ plotted as a
  function of $\lambda t$. The set of three thin red curves decaying at short
  times correspond to $\a_{1} = \lambda, \a_{2}=0, \mu/\lambda = 0.01,
  \nu/\lambda = 0.99$, while the three black curves decaying at longer times
  correspond to $\mu/\lambda = \nu/\lambda = 0.5$. The effects of
  increased chain length $L$ are more dramatic when particle decay is
  faster and length-dependent stochastic tunneling becomes
  rate-limiting. (b) Plots of $\ln(\lambda T)$ as a function of $L$
  for $\a_{1}/\lambda =1,10,10^{4}$ with fixed $\a_{2}=0, \mu/\lambda
  = 0.1$, and $\nu/\lambda = 0.9$. The two dashed curves correspond to
  the asymptotic limits in Eq.~\ref{TNAPPROX}.}
\label{NN_LIN}
\end{center}
\end{figure}

If there is death ($\mu > 0$) but also immigration ($\alpha_{1}$
and/or $\alpha_{2} > 0$), the explicit expression for the overall
survival probability $\Sigma(t)$ can be found by using Eq.~\ref{SS} in
Eq.~\ref{SIGMAEXP}. In the constant $\lambda = \mu+\nu$ case, we find

\begin{equation}
\begin{array}{rl}\label{SIGMALINEAR}
\Sigma(t) &  \displaystyle = \exp\left[-\a_{1}\int_{0}^{t}(1-S_{1}(t'))\dd
  t'\right]\exp\left[-\a_{2}\int_{0}^{t}(1-S_{1}^{2}(t'))\dd
  t'\right] \\[13pt]
%
%
\: & = \displaystyle \exp\left[-{(\a_{1}+2\a_{2})\over \lambda}\left(\nu\over \lambda\right)^{L}
  \left( \lambda t - L-\lambda t{\Gamma(L,\lambda t)\over \Gamma(L)}+
       {\Gamma(L+1,\lambda t)\over \Gamma(L)}\right)\right] \\[13pt]
\: & \displaystyle \hspace{5mm}\times
\exp\left[-2{\a_{2}\over \lambda}\left({\nu\over \lambda}\right)^{2L}\left(L +\lambda t
  {\Gamma(L,\lambda t)\over \Gamma(L)}-{\Gamma(L+1,\lambda t)\over
    \Gamma(L)}-{\lambda t\over 2}\right)\right]\exp\left[\a_{2}\left({\nu
    \over \lambda}\right)^{2L} \int_{0}^{t}\left({\Gamma(L,\lambda
    t')\over \Gamma(L)}\right)^{2}\dd t'\right].
\end{array}
\end{equation}
%
When $\a_{2}=0$ (no double-particle immigration), the integral $T =
\int_{0}^{\infty} \Sigma(t)\dd t$ can be approximated in the small and
large limits of $\Omega \equiv (\a_{1}/\lambda)(\nu/\lambda)^{L}$ by
considering the structure of integrand $\Sigma(t)$ in
Eq.~\ref{T}\cite{BENDER}:

\begin{equation}
T \approx
\begin{cases}
\displaystyle {L\over \lambda}\left[1+{1\over \Omega L}\right],
& \displaystyle \Omega \equiv {\a_{1}\over \lambda}
\left({\nu \over \lambda}\right)^{L} \ll 1 
\\[13pt]
\displaystyle {1\over \lambda}\Gamma\left({L+2\over L+1}\right)\left[{(L+1)!\over 
\Omega L}\right]^{{1\over L+1}}\!\!\!, & \displaystyle 
{\Omega \over L!}\gg 1.
\end{cases}
\label{TNAPPROX}
\end{equation}
Fig.~\ref{NN_LIN}(a) shows exact survival probabilities of the
homogeneous sequential linear process for different values of chain
length $L$. For comparison, we plot curves corresponding to different
rate parameters $\mu$ and $\nu$ relative to the total uniform
transition rate $\lambda = \mu +\nu$.
Fig.~\ref{NN_LIN}(b) plots $\ln \lambda T$ as a function of chain
length $L$. For large $\Omega L$, the rate limiting step is
immigration and the MFPT is approximately the inter-immigration time,
normalized by the probability each immigration event eventually leads
to fixation (the $\Omega \ll 1$ limit in Eq.~\ref{TNAPPROX}).

%

\subsection{Nonlinear processes}

Now, consider cell replication processes where $p_{k}+q_{k}+r_{k} >
0$. When these higher order cellular processes arise, Eq.~\ref{SK} is
nonlinear for $N>1$, and the evaluation of survival probabilities
and first passage times must be approximated or computed numerically.
From Eq.~\ref{SIGMAEXP}, we can see that for sufficiently small
$\alpha_{1}/\lambda$, the survival probability will scale as
$\Sigma(t) \sim e^{-\a_{1}(1-\bar{S}_{1})t}$.  Note that if $\mu_{k} =
0$ for all $k$, the only steady-state solution to Eq.~\ref{SK} is
$S_{k}(t\to\infty)\equiv \bar{S}_{k}=0$.  Hence, $\Sigma(t)\sim
e^{-\a_{1}t}$, indicating that immigration is the rate limiting
step. In the following we we will provide results to a few specific
illustrative cases.

\subsubsection{Mean field Approximation}

The simplest approximation to the survival probability can be obtained
without using Eqs.~\ref{MATRIXS} and \ref{SIGMAEXP}.  The time rate of
change of survival is simply defined as the total probability flux
into absorbing states, {\it conditioned} on no particle having yet
entered any absorbing state \cite{FPTREVIEW}. In our problem, the {\it
  unconditioned} instantaneous particle flux into state $L+1$ is
$J_{\rm mf}(t) \equiv (p_{L}+q_{L}+\nu_{L})\langle n_{L}(t)\rangle$,
where $\langle n_{L}(t)\rangle$ is the expected occupation of state
$L$.  If we assume that the mean occupation is uncorrelated with the
probability $\Sigma(t)$ of survival, $\dot{\Sigma}_{\rm mf} \approx
-J_{\rm mf}(t)\Sigma_{\rm mf}$.  This approximation is exact when
particles are always independent and is widely used.  The survival
probability under this mean-field assumption is thus

\begin{equation}
\Sigma_{\rm mf}(t) =
\exp\left[-(p_{L}+q_{L}+\nu_{L})\int_{0}^{t}\langle n_{L}(t')\rangle
  \dd t'\right].
\end{equation}
The unconditioned occupation $\langle n_{L}(t')\rangle$ can be found
using mass-action equations for the particle density at each site. The
Laplace-transformed expected particle number can be written as

\begin{equation}
\langle \tilde{n}_{L}(s)\rangle = {(\a_{1} +2\a_{2})\over a_{L}
  s}\prod_{i=1}^{L} {a_{i} \over s+b_{i}},
\label{NNLAPLACE}
\end{equation}
where $a_{i} \equiv 2p_{i}+q_{i}+\nu_{i}$ and $b_{i} \equiv
\mu_{i}+\nu_{i}+p_{i}-r_{i}$. Like Eq.~\ref{SS},
this result shows that the mean-field survival probability of a
system injected at the first site is independent of the
specific order of the rates.  Moreover, upon
comparing Eq.~\ref{NNLAPLACE} to Eq.~\ref{SS}, we see that the mean
field survival probability $\Sigma_{\rm mf}(t) = \Sigma(t)$ is exact
if $\a_{2} = p_{i} = q_{i} = r_{i} = 0$.

For general rates but uniform $a_{i} = a$ and $b_{i}=b$, the general mean-field
approximation for the survival probability is

\begin{equation}
\Sigma_{\rm mf}(t) = \exp\left[-{(\a_{1}+2\a_{2})\over
    b}{(p+q+\nu)a^{L-1}\over b^{L}}\left(bt - L -bt{\Gamma(L,bt)\over
    \Gamma(L)}+{\Gamma(L+1,bt)\over \Gamma(L)}\right)\right],
\label{SIGMAMF}
\end{equation}
which has a form analogous to Eq.~\ref{SIGMALINEAR}.  To explicitly
see that $\Sigma_{\rm mf}(t)$ is not exact when any $\a_{2}, p, q, r
>0$, consider the single intermediate state case $L=1$.  In this case,
Eq.~\ref{SK} can be solved exactly to yield explicit expressions for
$S_{1}(t)$ and $\Sigma(t)$:

\begin{equation}
\Sigma(t) = e^{-\alpha_{1}(1-S_{-})t}e^{-\alpha_{2}(1-S_{-}^{2})t}
\left[{(S_{+}-S_{-})e^{\gamma t}\over (S_{+}-1)e^{\gamma t} +
    (1-S_{-})}\right]^{\alpha_{1}/r +\alpha_{2}\lambda/r^{2}}
\exp\left[-{\alpha_{2} \over r}{(S_{+}-1)(1-S_{-})(e^{\gamma t}
    -1)\over (S_{+}-1)e^{\gamma t} +(1-S_{-})}\right],
\label{SIGMAN1}
\end{equation}
%
where 

\begin{equation}
S_{\pm} = {\lambda\over 2r} \pm {\lambda\over 2r}\sqrt{1-4\mu r/\lambda^{2}}
\end{equation}
and 

\begin{equation}
\gamma = (S_{+}-S_{-})r = \lambda \sqrt{1-4\mu r/\lambda^{2}}.
\end{equation}

\begin{figure}[t]
\begin{center}
\includegraphics[width=3.3in]{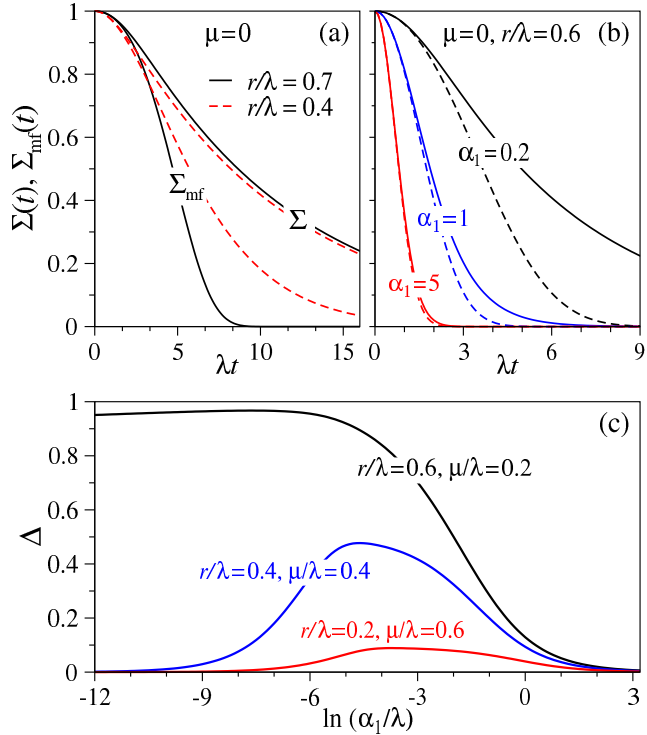}
\caption{Comparison between exact solutions and mean-field
  approximations for $L=1$. (a) $\Sigma(t)$ and $\Sigma_{\rm mf}(t)$
  for $\mu=0$ and $r/\lambda =0.7$ (solid) and $r/\lambda =0.4$
  (dashed). As expected, differences are larger for larger values of
  $r/\lambda$, where survival probability and the mean occupation
  $\langle n_{1}(t)\rangle$ share more correlations. (b) $\Sigma(t)$
  (solid) and $\Sigma_{\rm mf}(t)$ (dashed) for different values of
  single-particle immigration $\a_{1}$ and fixed $\a_{2}=\mu=0,
  r/\lambda = 0.6$, and $(p+q+\nu)/\lambda=0.4$. The difference is
  largest for smaller $\a_{1}$ where immigration is rate limiting, and
  the first arrival at the absorbing state $k=2$ is more likely from
  particles that have replicated at $k=1$. (c) Relative errors of
  MFPTs $\Delta \equiv (T - T_{\rm mf})/T$ as a function of $\a_{1}$
  for the combinations of $r/\lambda, \mu/\lambda$ indicated. When
  rates of nonlinear processes ($r$ in this case) are large, the error
  is large. In the limit of vanishing $r/\lambda$, mean-field theory
  becomes exact and $\Delta$ vanishes.}
\label{N1}
\end{center}
\end{figure}

Fig.~\ref{N1} explicitly shows the difference 
between $\Sigma(t)$ and $\Sigma_{\rm mf}(t)$ (Eq.~\ref{SIGMAMF})
%
%
for various values of $\a_{2}, p, q, r > 0$.
The discrepancy between the exact and mean-field results vanishes as
$(\a_{1}/b)(a/b)^{L}/L! \gg 1$. In this limit, the numbers of
particles derived from independently immigrated lineages are
sufficiently large such that the effects of correlations among their
branching times are small.  The mean-field limit can also be derived
by considering the solution to $S_{1}$ in the short time limit when it
deviates only slightly from unity.  Linearization of Eq.~\ref{SK}
about $S_{k}= 1$ results in a set of equations whose solution also
yield the mean-field result of Eq.~\ref{SIGMAMF}.

\subsubsection{Numerical results}

To investigate the effects of nonlinear proliferative processes on
evolution and first passage times in larger systems, we
solve Eq.~\ref{SK} numerically and use Eqs.~\ref{SIGMAEXP} and \ref{T}
to find survival probabilities and MFPTs. Since Eq.~\ref{SK} is
nonlinear, we expect the ordering of the rates and positioning of
defects along the chain to influence first passage times, in
contradistinction to linear processes in which spatial ordering of
rates does not play a role.

\begin{figure}[t]
\begin{center}
\includegraphics[width=3.6in]{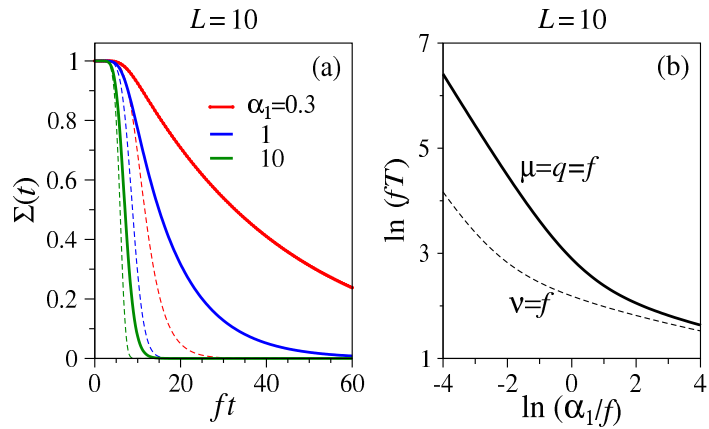}
\caption{Comparison of two mean-field-equivalent processes $\nu=f$ and
  $\mu=q=f$ along a chain of length $L=10$.  (a) The dashed curves are
  $\Sigma(t)$ for the linear process $\nu=f$ and $\mu=p=q=r=0$, while
  the thick solid curves correspond to numerical solutions of
  Eq.~\ref{SK} for the nonlinear process $\mu=q=f$ and
  $\nu=p=r=0$. Due to the independent decay processes, the MFPT of the
  nonlinear process is always greater than that of the linear
  process. (b) MFPTs as functions of $\ln(\a_{1}/f)$. Despite the
  mean-field equivalence, mean-field approximations to the FPTs are
  qualitatively inaccurate.}
\label{MORAN}
\end{center}
\end{figure}

We first compare proliferative processes with an irreversible-mutation
linear Moran-type process in which asymmetric differentiation occurs
followed immediately by death of the parent cell. This assumption is
typically used to enforce fixed population (in the absence of
immigration) and in our framework corresponds to $\nu_{k} > 0$ and
$\mu_{k}=p_{k}=q_{k}=r_{k}=0$.  This process is linear and a
mean-field assumption yields exact results.  A related nonlinear
process can be defined by $q_{k} = \mu_{k}>0$ (and
$\nu_{k}=p_{k}=r_{k}=0$). This process will give rise to identical
expected populations $\langle n_{k}(t)\rangle$ if $q_{k}$ are assigned
the same values as $\nu_{k}$ used in the linear Moran-type process.
Here, asymmetric differentiation and death are balanced such that the
mean occupations are identical to those derived from the linear
process $\mu_{k}=p_{k}=q_{k}=r_{k}=0$.  However, in the linear
process, mutation and death of the parent particle are completely
correlated, unlike in the nonlinear process ($q_{k}= \mu_{k} > 0$) in
which they occur independently. The nonlinear process allows
fluctuations in the total population to affect FPT statistics.  In
Fig.~\ref{MORAN}, $\Sigma(t)$ and the MFPTs between two processes with
uniform intrinsic rate $f$, ($\nu=f$, $p=q=r=\mu=0$) and ($q=\mu=f$,
$\nu=p=r=0$), are contrasted.
The results in Fig.~\ref{MORAN} can also be qualitatively understood
from the likelihood of any particle at site $k$ generating one at site
$k+1$. If $\mu = q = f > 0$, then any single cell would have a
probability of only one half of generating an advancing daughter cell
particle.  However, in the linear Moran-type process with $\nu = f$,
all particles will eventually move forward.

In the small $\a_{1}/f$ limit, the MFPT of the nonlinear proliferative
process scales as $T \sim
\left[\a_{1}(1-\bar{S}_{1})\right]^{-1}$. For $\mu = q = f$,
$\bar{S}_{1} = L/(L+1)$, and $T \sim (L+1)/\a_{1} > T_{\rm mf}$, where
$T_{\rm mf}$ is the exact mean-field result for the MFPT of the linear
Moran-like process, which can be found from Eq.~\ref{TNAPPROX} or by
using Eq.~\ref{SIGMAMF} in Eq.~\ref{T}. When $\a_{1}/f$ is large, the
number of statistically independent particles in the system is large
and the survival probability of the proliferative process will
approach a common mean-field limit (Eq.~\ref{SIGMAMF}).  Thus, the
relative difference between the MFPTs of the linear spontaneous
mutation process and the mean-field-equivalent nonlinear process
diminishes at large injection rates $\a_{1}$ (and $\a_{2}$).
Nonetheless, cells in the proliferative process have a nonzero death
rate and the MFPT is bounded above by that of the linear process.
Therefore, in terms of reaching the absorbing state, we observe that
the linear irreversible Moran-type process is always faster.

Next, consider another proliferative process that might be expected
to yield similar FPTs as the linear Moran-like process. If cells undergo only
symmetric differentiation and death with rates $p=\mu =f$ and
$q=r=\nu=0$, a parent cell can die or beget two differentiated
daughters that each die at the same rate. Even though the expected
populations of this process and of the irreversible Moran-type process
($\nu = f$) differ, the mean positions of the lead particle are
equal (conditioned on survival).
\begin{figure}[t]
\begin{center}
\includegraphics[width=3.6in]{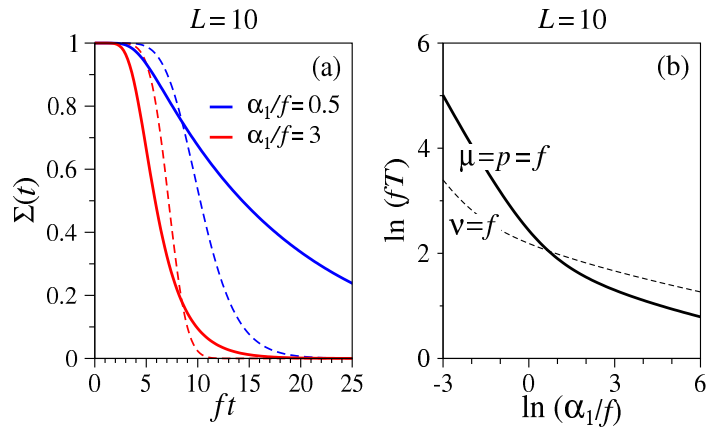}
\caption{Comparison of two processes with similar mean lead-particle
  positions.  (a) This dashed curves are $\Sigma(t)$ (which is
  equivalent to $\Sigma_{\rm mf}$) for the linear process $\nu=f$ and
  $\mu = p=q=r=0$, while the thick solid curves correspond to
  numerical solutions of Eqs.~\ref{SK} and \ref{SIGMAEXP} for the
  nonlinear process $\mu=q=f$ and $\nu=p=r=0$. Due to independent
  decay process, the MFPT of the nonlinear process is always greater
  than that of the linear process. (b) The MFPTs as functions of
  $\ln(\a_{1}/f)$ for these two processes also dramatically differ,
  but a cross-over occurs.}
\label{MORANP}
\end{center}
\end{figure}
Fig.~\ref{MORANP}(a) shows the survival probabilities of the two
processes for two different values of immigration.  For small
immigration rates $\a_{1}/f$, the linear (mean-field) process reaches
the absorbing state faster, while for high immigration rates, the
proliferative process is faster. Fig.~\ref{MORANP}(b) plots the MFPT
of the two processes as a function of injection rate.  For small
$\a_{1}/f$, the exact MFPT $T_{\rm mf}$ of the linear process can
again be found from the first limit in Eq.~\ref{TNAPPROX}, while the
MFPT of the nonlinear proliferative process scales as $T \approx
\left[\a_{1}(1-\bar{S}_{1})\right]^{-1}$. In this case, the lineage
associated with each injected cell has a possibility of becoming
extinct before fixation, resulting in a MFPT diverging as $1/\a_{1}$.
For $L=10$, $\bar{S}_{1} \approx 0.861$ and $T \approx
(0.139\a_{1})^{-1} > T_{\rm mf}$.  When $\a_{1}/f$ is large,
$\Sigma(t)$ for the nonlinear proliferative process approaches the
mean-field result in Eq.~\ref{SIGMAMF}. Moreover, the associated MFPT
can be shown to be less than the MFPT for the linear process.  Thus,
there is a cross-over at a particular value of immigration below which
the linear process becomes evolutionarily faster than the
proliferative process. For large $\a_{1}$, immigration is sufficiently
fast to allow overall proliferation to push lead particles to overtake
those of the corresponding linear Moran-type process, leading to a
smaller MFPT.

Finally, we illustrate the effects of two types of deserts (or
bottlenecks) and two types of oases in an otherwise uniform
evolutionary chain. Bottlenecks or deserts at site $L^{*}$ may arise
from an enhanced death rate $\mu^{*}$, or from a suppression in 
$\nu^{*}$, $p^{*}$, and/or $q^{*}$.  A local
oasis can modeled by increased proliferation rates such as $r^{*}$ or
$p^{*}$.  For example, Fig.~\ref{ZRP} depicts a sequential process
with an enhanced growth rate at site $L^{*}$.
\begin{figure}[t]
\begin{center}
\includegraphics[width=3.6in]{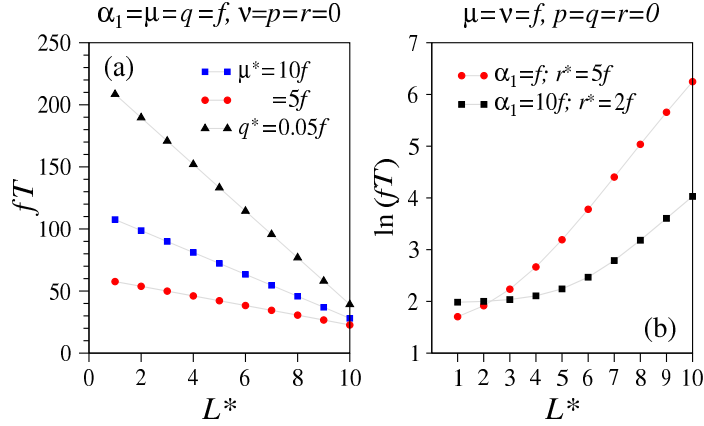}
\caption{Spatial dependence of bottlenecks and oases. (a) Dependence
  of the MFPT on the position of a bottleneck. In an otherwise uniform
  chain with $q=\mu =f$, $\nu =p=r=0$, cells at site $L^{*}$ die with
  increased rates $\mu_{L^{*}} \equiv \mu^{*} = 5f$ (red circles) and
  $\mu^{*} = 10f$ (blue squares).  Alternatively, this bottleneck site
  may have a diminished asymmetric division rate $q_{L^{*}}\equiv
  q^{*} =0.05f$ (black triangles). (b) The dependence of the MFPT on
  the position of an oasis at a site with no death ($\mu_{L^{*}}=0$),
  enhanced growth rates $r_{L^{*}}\equiv r^{*} = 2f, 5f$, and
  corresponding immigration rates $\a_{1} = f$ and $\a_{1}=10f$,
  respectively. In the rest of the chain, there are no proliferative
  processes $(p=q=r=0)$ and cells both spontaneously mutate and die
  with rate $\nu=\mu=f$.}
\label{BOTTLENECK}
\end{center}
\end{figure}
Fig.~\ref{BOTTLENECK} plots the MFPT for a bottleneck (a), and an
oasis (b), at different positions along the chain. For the parameters
used, bottlenecks are most effective at slowing down fixation when
placed near the start the chain; conversely, an oasis is most
effective at speeding up fixation when placed near the start of the
chain.  

The linear dependence on bottleneck position shown in
Fig.~\ref{BOTTLENECK}(a) can be understood by viewing this scenario as
a FPT problem in the second segment of the chain $L^{*} < \ell \leq
L+1$.  Related sequential segmentation methods have also been used to
self-consistently compute steady-state transport fluxes across
excluding 1D lattices \cite{KOLO1998,LAKATOS04}. Here, the bottleneck
reduces the effective immigration rate into the second segment. If the
bottleneck is sufficiently strong (as are the cases shown in
Fig.~\ref{BOTTLENECK}(a)), immigration into the second segment is
rate-limiting and since $\nu=0$, we expect the MFPT to scale as
$1/(L-L^{*}+1)$.

The effect of an oasis site in the presence of an otherwise uniform
process involving death and spontaneous mutation is to decrease the
MFPT, as shown in Fig.~\ref{BOTTLENECK}(b). If the rates at site
$L^{*}$ are such that $r^{*} > \mu^{*}+\nu^{*}$, there can be
unlimited growth and the rate of immigration into site $L^{*}+1$ will
exponentially increase time. Thus, an oasis near the beginning of the
evolutionary chain will strongly drive immigration into the remaining
segment and be more effective at reducing the MFPT to fixation
compared to one that is hard to get to near the end of the chain.

An oasis with a positive net growth rate leads to an unbounded
population at long times.  However, our approach does not allow for
interactions and constraints such as carrying capacity. Nonetheless,
if the first arrival times to $L+1$ are much smaller than the time it
takes for any site to reach carrying capacity ($K\gg
\exp\left[(r^{*}-\mu^{*})T\right]$), our unlimited growth model
still provides a reasonable approximation to the FPT.

\begin{figure}[h!]
\begin{center}
\includegraphics[width=3.6in]{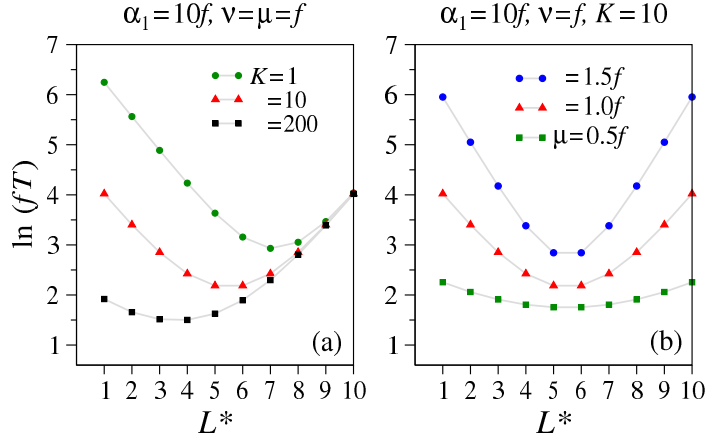}
\caption{MFPT in the presence of an oasis with large growth rate
  $r^{*}\to \infty$ but a finite carrying capacity $K$.  (a) $\ln
  f T$ for various carrying capacities $K$ at fixed
  immigration rate $\a_{1} = 10f$ and spontaneous mutation and death rate
  $\nu=\mu = f$. (b) When $\a_{1}= K\nu = 10f$, both effective
  immigration rates are equal and the MFPT-minimizing position $L^{*}
  \approx L/2$ (Eq.~\ref{NSTAR_MIN}).}
\label{NSTAR}
\end{center}
\end{figure}

In the opposite limit of small carrying capacity ($K\ll
\exp\left[(r^{*}-\mu^{*})T\right]$) another approximation to the MFPT
can be obtained.  We can model an oasis by assuming that in an
otherwise homogeneous chain along which $p=q=r=0$, site $L^{*}$
carries a growth process with a carrying capacity $K$ and $r^{*} \to
r^{*}(1-n_{L^{*}}/K)$. We also assume that $\mu^{*}=0$ and that
$r^{*}$ is greater than all other rates in the model. Therefore, once
the first particle arrives at site $L^{*}$, its population quickly
rises to a level $\sim K$. These cells then feed into site $L^{*}+1$
through mutational processes described by $\nu, p,$ or $q$. By
considering two linear processes joined by an oasis at site $L^{*}$,
the MFPT to state $L+1$ can be approximated as the mean time to reach
$L^{*}$ plus the time to reach state $L+1$ given an effective
immigration rate $K\nu$ into site $L^{*}+1$.  Not only does the MFPT
depend on the spatial structure of the inhomogeneity, but in many
cases, there will be an optimal placement of an oasis which most
effectively reduces the overall MFPT. Such an optimal placement can be
explicitly seen by considering Eq.~\ref{TNAPPROX} in the small
immigration limit:

\begin{equation}
\begin{array}{rl}T(L,L^{*}) & \approx T(L^{*}-1) + 
T(L-L^{*}) \\[13pt] 
\: & \displaystyle \approx {1\over \Omega \lambda} + 
{1\over
  \Omega^{*}\lambda}+ {L\over \lambda},
%
%
\end{array}
\end{equation}
where $\Omega \equiv (\a_{1}/\lambda)(\nu/\lambda)^{L^{*}-1}$ and
$\Omega^{*} \equiv (K\nu/\lambda)(\nu/\lambda)^{L-L^{*}}$.  This
approximation clearly shows a position-dependent MFPT provided
$\nu/\lambda < 1$ ($\mu>0$).  The position $L^{*}_{\rm min}$
which yields the smallest MFPT in the $\Omega L^{*},
\Omega^{*}(L-L^{*}) \ll 1$ limit can be approximated by solving
$\partial T(L,L^{*})/\partial L^{*} = 0$:

\begin{equation}
L^{*}_{\rm min} \approx {\ln \left({\a_{1}\over K\nu}\right)\over 2\ln
  \left({\lambda \over \nu}\right)} +{L+1\over 2},
\label{NSTAR_MIN}
\end{equation}
which shows that when $K\nu\approx \a_{1}$, the oasis lowers the
MFPT the most when placed near the midpoint of the chain. 
Eq.~\ref{NSTAR_MIN} provides good estimates of the optimal 
oasis position $L^{*}_{\rm min}$ and its dependences on
rates.

In Fig.~\ref{NSTAR}(a) we use Eqs.~\ref{SIGMAEXP} and \ref{T} to
compute the MFPT of a two-segment chain. For the segment before the
oasis, we use $T(L^{*}-1) = \int_{0}^{\infty}\Sigma(t;\a_{1}=f,
\mu=\nu=f, L^{*}-1) \dd t$, while for the second segment, $T(L-L^{*})
= \int_{0}^{\infty}\Sigma(t;\a_{1}\equiv K\nu, \mu=\nu=f, L-L^{*}) \dd
t$. Evaluating the total MFPT $T(L^{*}-1)+T(L-L^{*})$ clearly shows
that the most effective positioning of an oasis is such that the
segment with rate-limiting immigration is shortest.  Since changes in
$\mu$ only affect $L^{*}_{\rm min}$ logarithmically, small changes in
the death rate do not affect the optimal oasis position. However, when
$\mu$ increases, as shown in Fig.~\ref{NSTAR}(b), the MFPTs across
each segment increases exponentially with its length, increasing the
sensitivity of the overall MFPT to $L^{*}$.

\section{Discussion \& Conclusions}


We have formulated an efficient way to analyze FPTs on a network
containing multiple, mutating, and proliferating particles. Our model
allows one to naturally study stochastic evolutionary processes and
explicitly include cell fate decisions, fluctuations in total number,
and immigration. A number of asymptotic limits are explored and
comparisons with mean-field calculations of survival probabilities
performed. Kinetic Monte Carlo simulations were also performed and
checked against our results.  Our main findings illustrate the
importance of specific cellular transitions and how mean-field
assumptions can be misleading when used to compute first arrival times.
%
%
Even though expected particle numbers of a noninteracting particle
system can typically be found exactly using mean-field approximations,
our results explicitly show how survival probabilities and first
passage time statistics cannot be treated using simple mean-field
approximations if particles can proliferate.  These discrepancies are
prominent in conditions of low populations, as encountered in
stochastic tunneling.



Furthermore, proliferative processes, including symmetric and
asymmetric cell differentiation, render FPTs dependent on the order of
the transition rates along a sequential evolutionary chain. For many
scenarios, we find bottlenecks are most effective at increasing the
MFPT when placed at the beginning of an evolutionary chain, while an
unlimited oasis reduces the MFPT most effectively at the beginning of
the chain.  If the growth rate of an oasis site is faster than any
other time scale, the mean times to the terminal state can be
approximated by the mean time for the first cell to arrive at the
oasis, plus the time for the progeny of any cell arising from an oasis
to arrive at the terminal site. In the presence of regulating
interactions that generate {\it e.g.,} a carrying capacity $K$, we
find intermediate oasis positions that optimally reduce the MFPT to
the final $L+1$-state. This optimal position is qualitatively
determined by the ratio of the effective immigration rates into each
of the segments and deviates from the halfway point by the log of the
ratio of immigration rates, with the shorter segment associated with
the smaller effective immigration rate.


Collectively, our results suggest that fixation times across a number
of biological systems may be sensitive to the precise transitions
allowed.  Examples include stem cell differentiation
\cite{ROSHAN2014} and mutation \cite{MCHALE2014}, where each
differentiation or mutational state is represented by distinct nodes.
Our approach is also particularly appropriate for modeling progression
and drug resistance in cancer.  Since mutated or precancerous cells
may likely have only a small fitness advantage
\cite{BEERENWINKEL2007}, the numbers of cells in these states may be
small, and the effects of proliferative nonlinearity may be important.
In such cases, cell states that are drug resistant will do the most
harm when occurring at the beginning, or in the interior of the
mutational sequence, depending on, respectively, whether a carrying
capacity arises or not. We have investigated only simple, irreversible
transitions along a 1D sequential chain. Extensions to more complex
networks and nonexponentially distributed processes (such as
cell-cycle timing) can be readily investigated by numerically solving
Eqs.~\ref{SJ0} and \ref{SIGMA0}. More complex distributions of
different transition rates can also be easily treated numerically.



\section{Acknowledgements}

TC was supported by the NSF through grant DMS-1021818, the Army
Research Office through grant 58386MA, and the DoD through grant
W911NF-13-1-0117. YW was supported through the Cross-disciplinary
Scholars in Science and Technology (CSST) program at UCLA. The authors
also wish to acknowledge the support of the KITP at UCSB through NSF
PHY11-25915.

\bibliography{oasis11_jtb.bbl}

\end{document}